\def\aj{{AJ}}

\def\apj{{ApJ}}

\def\mnras{{MNRAS}}

\documentclass[cup5b]{caps}
\usepackage{graphicx}
\usepackage{amssymb}
\usepackage{ociwsymp1}
\HeadText{Z. Haiman}

\def\plotone#1{\centering \leavevmode
\includegraphics[width=.95\columnwidth]{#1}}

\def\gsim{\;\rlap{\lower 2.5pt
 \hbox{$\sim$}}\raise 1.5pt\hbox{$>$}\;}
\def\lsim{\;\rlap{\lower 2.5pt
   \hbox{$\sim$}}\raise 1.5pt\hbox{$<$}\;}
\def\ga{\;\rlap{\lower 2.5pt
 \hbox{$\sim$}}\raise 1.5pt\hbox{$>$}\;}
\def\la{\;\rlap{\lower 2.5pt
   \hbox{$\sim$}}\raise 1.5pt\hbox{$<$}\;}
\def\HH{H$_2$}

\newcommand{\lya}{Ly$\alpha$ }
\newcommand\msun{M_\odot}
\newcommand{\Htwo}{{\rm H_2}}

\begin{document}

\pagenumbering{arabic}

\author[]{ZOLT\'AN HAIMAN\\Department of Astronomy, Columbia University}

\chapter{The First Nonlinear Structures \\ 
         and the Reionization History of \\ 
         the Universe}

\begin{abstract}
In cosmological models favored by current observations, the first
astrophysical objects formed in dark matter halos at redshifts
starting at $z\gsim 20$, and their properties were determined by
primordial ${\rm H_2}$ molecular chemistry. These protogalaxies were
very abundant, but substantially less massive than typical galaxies in
the local Universe.  Extreme metal-poor stars, and massive black
holes in their nuclei reionized the bulk of the hydrogen in the
intergalactic medium.  Reionization may have taken place over an
extended redshift interval, ending around $z\approx 7$.  Observational
probes of the process of reionization may soon be afforded by studying
the polarization of the cosmic microwave background 
anisotropies, as well as by studying the spectra and abundance of
distant Ly$\alpha$-emitting galaxies. Here we review theoretical
expectations on how and when the first galaxies formed, and summarize
future observational prospects of probing hydrogen reionization.
\end{abstract}

\section{Introduction}
\label{sec:introduction}

Recent measurements of the cosmic microwave background (CMB)
temperature anisotropies, determinations of the luminosity distance to
distant Type Ia supernovae, and other observations have led to
the emergence of a robust ``best-fit'' cosmological model with energy
densities in cold dark matter (CDM) and ``dark energy'' of
$(\Omega_{\rm m},\Omega_{\rm \Lambda})\approx (0.3,0.7)$ (see Bahcall
et al. 1999 for a review, and references therein).
The growth of density fluctuations and their evolution into
nonlinear dark matter structures can be followed in this
cosmological model in detail from first principles by semi-analytic
methods (Press \& Schechter 1974; Sheth, Mo, \& Tormen 2001).  More recently,
it has become possible to derive accurate dark matter halo mass
functions directly in large cosmological $N$-body simulations (Jenkins
et al. 2001).  Structure formation in a CDM-dominated Universe is
``bottom-up,'' with low-mass halos condensing first.  Dark matter
halos with the masses of globular clusters, $10^{5-6}\msun$, are
predicted to have condensed from $\sim 3\sigma$ peaks of the initial
primordial density field as early as $\sim$1\% of the current age of
the Universe, or redshift $z\approx  25$.

It is natural to identify these condensations as the sites where the
first astrophysical objects, such as stars, or quasars, were born.
The nature of the objects that form in these early dark matter halos
is currently one of the most rapidly evolving research topics in
cosmology. Progress is being driven by recent observational and
theoretical advances, and also by the promise of next generation
instruments in several wavelength bands, such as the {\it James Webb
Space Telescope (JWST)} in the infrared, the {\it Low Frequency Array
(LoFAr)} in the radio, {\it XMM-Newton} in the X-rays, and {\it Laser
Interferometer Space Antenna (LISA)} in gravity waves.

A comprehensive review of the status of the field two years ago was
provided by Barkana \& Loeb (2001); a more focused review on the role
of ${\rm H_2}$ molecules at high redshift was given by Abel \& Haiman
(2001). In the present paper, we briefly summarize the main
theoretical issues in high-redshift structure formation, and then
focus on progress in the last two years, both in theory and in
observation. It is appropriate to single out\footnote{As this article
went to press, the first results by the {\it Wilkinson Microwave
Anisotropy Probe} ({\it WMAP}; Bennett et al. 2003) experiment were 
announced.  See the
Appendix for a summary of the implications.} the recent discovery of a
Gunn-Peterson (1965, hereafter GP) trough in the spectra of a few high-redshift
quasars (Becker et al. 2001; Fan et al. 2003). The absence of any
detectable flux shortward of $\sim(1+z)1216$ \AA\ in the spectra of
these $z>6$ sources has raised the tantalizing possibility that at
these redshifts we are directly probing into the epoch
reionization. It has also brought into sharp focus the question of how
to distinguish observationally various reionization histories.  We
will critically discuss these issues below.

\section{Theoretical Expectations}

\subsection{The First Galaxy: When and How Massive?}

Baryonic gas that falls into the earliest nonlinear dark matter halos
is shock heated to the characteristic virial temperatures of a few
hundred K. It has long been pointed out (Rees \& Ostriker 1977;
White \& Rees 1978) that such gas needs to lose its thermal energy in
order to continue contracting, or in order to fragment --- in the
absence of any dissipation, it would simply reach hydrostatic
equilibrium, and would eventually be incorporated into a more massive
halo further down the halo-merger hierarchy.  While the formation of
nonlinear dark matter halos can be followed from first principles,
the cooling and contraction of the baryons, and the ultimate formation
of stars or black holes (BHs) in these halos, is much more difficult
to model {\it ab initio}. Nevertheless, it is useful to identify four
important mass scales, which collapse at successively smaller
redshifts:
%\begin{enumerate}
%\item 
(1) gas contracts together with the dark matter only in dark halos
above the cosmological Jeans mass, $M_{\rm J}\approx 10^4
[(1+z)/11]^{3/2}\,\msun$, in which the gravity of dark matter can overwhelm
thermal gas pressure;
%\item 
(2) gas that condensed into Jeans-unstable halos can cool and
contract further in halos with masses above $M_{\rm H_2} \gsim
10^5[(1+z)/11]^{-3/2}\,\msun$ (virial temperatures of $T_{\rm vir}\gsim
10^2$ K), provided there is a sufficient abundance of ${\rm H_2}$
molecules, with a relative number fraction at least $n_{\rm H_2}/n_{\rm
H}\approx  10^{-3}$);
%\item 
(3) in halos with masses above $M_{\rm H} \gsim
10^8[(1+z)/11]^{-3/2}\,\msun$ (virial temperatures of $T_{\rm vir}\gsim
10^4$ K), gas can cool and contract via excitation of atomic
Ly$\alpha$, even in the absence of any ${\rm H_2}$ molecules; and
%\item 
(4) in halos with masses above $M_{\rm H} \gsim
10^{10}[(1+z)/11]^{-3/2}\,\msun$ (virial temperatures of $T_{\rm
vir}\gsim 2\times10^5$ K), gas can cool and contract, even in the
face of an existing photoionizing background.
%\end{enumerate}

The first of these scales is obtained simply by balancing
gravitational and pressure forces. The second scale is obtained by
requiring efficient cooling via roto-vibrational levels of ${\rm H_2}$
molecules, on a time scale shorter than the age of the Universe at the
appropriate redshift.  The calculations of the appropriate cooling
functions for molecular hydrogen seem to be converging (Galli \& Palla
1998 has done the most recent computations; see Flower et al. 2001 for
a review and other references).  The third scale is obtained by
requiring efficient (Ly$\alpha$ line) cooling via atomic H. The fourth
scale is obtained in detailed spherical collapse calculations (Thoul
\& Weinberg 1996).

In the earliest, chemically pristine clouds, radiative cooling is
dominated by ${\rm H_2}$ molecules. As a result, gas-phase ${\rm H_2}$
``astro-chemistry'' is likely to determine the epoch when the first
astrophysical objects appear --- a conclusion reached already in the
pioneering works by Saslaw \& Zipoy (1967) and Peebles \& Dicke
(1968). Several papers constructed complete gas-phase reaction
networks and identified the two possible ways of gas-phase formation
of ${\rm H_2}$ via the ${\rm H_2^+}$ or ${\rm H^-}$ channels.  These
were applied to derive the ${\rm H_2}$ abundance under densities and
temperatures expected in collapsing high-redshift objects (Hirasawa
1969; Matsuda, Sato, \& Takeda 1969; Palla, Salpeter, \& Stahler 1983; 
Lepp \& Shull 1984;
Shapiro \& Kang 1987; Kang et al. 1990; Kang \& Shapiro 1992;
Shapiro, Giroux, \& Babul 1994).  Studies that incorporate ${\rm H_2}$
chemistry into cosmological models and address issues such as
nonequilibrium chemistry, dynamics, or radiative transfer, have
appeared only relatively more recently. Haiman, Thoul, \& Loeb (1996)
and Tegmark et al. (1997) studied the masses and redshifts of the
earliest objects that can collapse and cool via ${\rm H_2}$.  The
first three-dimensional (3D) cosmological simulations that incorporate
${\rm H_2}$ cooling date back to Gnedin \& Ostriker (1996) and Abel et
al. (1997).

The basic picture that emerged from these papers is as follows.  The
${\rm H_2}$ fraction after recombination in the smooth ``protogalactic''
gas is small ($x_{\rm H_2}=n_{\rm H_2}/n_{\rm H}\approx 10^{-6}$). At high
redshifts ($z\ga 100$), ${\rm H_2}$ formation is inhibited even in
overdense regions because the required intermediaries ${\rm H_2^+}$
and H$^-$ are dissociated by CMB
photons.  However, at lower redshifts, when the CMB energy density
drops, a sufficiently large ${\rm H_2}$ abundance builds up inside
collapsed clouds ($x_{\rm H_2}\approx 10^{-3}$) at redshifts $z\la 100$ to
cause cooling on a time scale shorter than the dynamical time.
Sufficient \HH\ formation and cooling is possible only if the gas
reaches temperatures in excess of $\sim 200$ K; or masses of few
$\sim 10^5[(1+z)/11]^{-3/2} \,{M_\odot}$.  The efficient gas cooling
in these halos suggests that the first nonlinear object in the
Universe was born inside a $\sim 10^5\,{M_\odot}$ dark matter halo
at redshift $z\approx 20$ (corresponding to a $\sim3\sigma$ peak of the
primordial density peak).

The nature of the first object is considerably more difficult to
elucidate. Nevertheless, the two most natural possibilities are for
stars or BHs, or perhaps both, to form. The behavior of
gas in a cosmological ``minihalo'' is a well-defined problem that has
recently been addressed in 3D numerical simulations
(Abel, Bryan, \& Norman 2000, 2002; Bromm, Coppi, \& Larson 1999, 2002).
These works have been able to follow the contraction of gas to much
higher densities than previous studies. They have shown convergence
toward a temperature/density regime of $T \approx 200\,~{\rm K}$ and 
$n \approx 10^{4} \, {\rm cm}^{-3}$, dictated by the critical
density at which the excited states of ${\rm H_2}$ reach equilibrium
and cooling becomes less efficient (Galli \& Palla 1998).  The 3D
simulations suggest that the mass of the gas fragments exceeds
$10^{2}-10^{3} \, {M_\odot}$, and, therefore that the first stars
in the Universe may have been unusually massive (but see Nakamura \&
Umemura 2002, who argue using 1D and 2D simulations that the initial
mass function may have been bimodal, with a second peak around $1-2\,
{M_\odot}$).  An important consequence of this conclusion is that
the earliest stars had an unusually hard spectrum --- because they were
metal free (Tumlinson \& Shull 2000) and also because they were
massive (Bromm, Kudritzki, \& Loeb 2001), possibly capable of ionizing
helium in addition to hydrogen.

\subsection{Radiative Feedback: Negative or Positive?}

The first objects will inevitably exert prompt and significant
feedback on subsequent structure formation.  This is because any soft
UV radiation produced below 13.6~eV and/or X-rays above $\ga 1$~keV
from the first sources can propagate across the smooth intergalactic
hydrogen gas, influencing the chemistry of distant regions (Dekel \&
Rees 1987).  Soft UV radiation is expected either from a star or an
accreting BH, with a BH possibly contributing X-rays
as well. Although recent studies find that metal-free stars have
unusually hard spectra, these do not extend to $\ga 1$~keV
(e.g., Tumlinson \& Shull 2000, but see also Glover \& Brandt 2003, who
find stellar X-rays to have a more significant effect).  The first
stars formed via ${\rm H_2}$ cooling are also expected to explode as
supernovae, also producing internal feedback within or near their own
parent cloud (Ferrara 1998; Omukai \& Nishi 1999).

External feedback from an early soft UV background were considered by
Haiman, Rees, \& Loeb (1997), Haiman, Abel, \& Rees (2000), Ciardi,
Ferrara, \& Abel (2000), and Machacek, Bryan, \& Abel (2001). It was
found that ${\rm H_2}$ molecules are fragile and are universally
photodissociated even by a feeble background flux (although Ricotti,
Gnedin, \& Shull 2002 find a positive effect: relatively near the
ionizing sources, ${\rm H_2}$ formation can be enhanced behind the H~II
ionization front in dense regions). ${\rm H_2}$ dissociation by the
$E<13.6$~eV photons occurs when the UV background is several orders of
magnitude lower than the value needed for cosmological reionization at
$z>5$ (and also than the level $\sim 10^{-21}~{\rm erg\, cm^{-2}\,
s^{-1}\, Hz^{-1}\, sr^{-1}}$ inferred from the proximity effect to exist at 
$z\approx 3$; Bajtlik, Duncan, \& Ostriker 1988).  The implication is a pause
in the cosmic star formation history: the buildup of the UV background  and the
epoch of reionization are delayed until larger halos ($T_{\rm vir}\ga
10^4$ K) collapse.  This is somewhat similar to the pause caused later
on at the hydrogen-reionization epoch, when the Jeans mass is
abruptly raised from $\sim 10^{4}~\,{M_\odot}$ to $\sim
10^{8-9}~\,{M_\odot}$.  An early background extending to the X-ray
regime would change this conclusion, because it catalyzes the
formation of ${\rm H_2}$ molecules in dense regions (Haiman, Rees, \&
Loeb 1996; Haiman et al. 2000; Glover \& Brandt 2003; but see
also Machacek, Bryan, \& Abel 2003 who find X-rays to have a less significant
effect).  If quasars with hard spectra ($\nu F_\nu \approx$ constant)
contributed significantly to the early cosmic background radiation
then the feedback might even be positive, and reionization can be
caused early on by minihalos with $T_{\rm vir}< 10^4$ K.

\subsection{The 2nd Generation: Atomic Cooling?}

Whether the first sources of light were massive stars or accreting
BHs (the latter termed ``miniquasars'' in Haiman, Madau, \& Loeb 1999)
is still an open question. Nevertheless, there is some tentative
evidence that reionization was caused by stars, rather than quasars:
high-redshift quasars appear to be rare, even at the impressive
depths reached by the optical Hubble Deep Fields (Haiman et al. 1999)
and the $>$1~Ms Chandra Deep Fields (Barger et al. 2003; see
also Mushotzky et al. 2000; Alexander et al. 2001; Hasinger 2003 for
earlier results on faint X-ray sources in the CDFs).  There also
seems to be a significant delay between the epochs of hydrogen and
helium reionizations, with helium ionized only at $z\approx 3$ (Songaila
1998; Heap et al. 2000), but hydrogen already at $z>6$ (see discussion
below), implying a relatively soft ionizing background spectrum.

If indeed the first light sources were stars, without emitting a
significant X-ray component above $\sim$1~keV, then efficient and
widespread star (and/or BH) formation, capable of reionizing the
Universe, had to await the collapse of halos with $T_{\rm vir} >
10^4$ K, or $M_{\rm halo} > 10^8 [(1+z)/11]^{-3/2} \, {M_\odot}$.
The evolution of such halos differs qualitatively from their less
massive counterparts (Oh \& Haiman 2002).  Efficient atomic line
radiation allows rapid cooling to $\sim 8000$ K; subsequently the gas
can contract to high densities nearly isothermally at this
temperature.  In the absence of ${\rm H_2}$ molecules, the gas would
likely settle into a locally stable disk, and only disks with unusually
low spin would be unstable.  However, the initial atomic line cooling
leaves a large, out-of-equilibrium residual free-electron fraction
(Shapiro \& Kang 1987; Oh \& Haiman 2002). This allows the molecular
fraction to build up to a universal value of $x_{\rm
H_2}\approx10^{-3}$, almost independently of initial density and
temperature (this is a nonequilibrium freeze-out value that can be
understood in terms of time scale arguments; see Susa et al. 1998 and  Oh
\& Haiman 2002).

Unlike in less massive halos, $\Htwo$ formation and cooling is much
less susceptible to feedback from external UV fields. This is because
the high densities that can be reached via atomic cooling. The ${\rm
H_2}$ abundance that can build up in the presence of a UV radiation
field $J_{21}$, and hence the temperature to which the gas will cool,
is controlled by the ratio $J_{21}/n$.  For example, in order for a
parcel of gas to cool down to a temperature of 500 K, this ratio has
to be less than $\sim 10^{-3}$ (where $J_{21}$ has units of $10^{-21}
{\rm erg \, s^{-1} \, cm^{-2} Hz^{-1} \, sr^{-1}}$, and $n$ has units
of ${\rm cm^{-3}}$).  Flux levels well below that required to fully
reionize the Universe strongly suppresses the cold gas fraction in
$T_{\rm vir} < 10^{4}$ K halos.  In comparison, Oh \& Haiman (2002)
showed that UV radiation with the same intensity has virtually no
impact on ${\rm H_2}$ formation and cooling in $T_{\rm vir} > 10^{4}$
K halos, where all of the gas is able to cool to $T=500$ K.  Indeed,
under realistic assumptions, the newly formed molecules in the dense
disk can cool the gas to $\sim 100$ K, and allow the gas to fragment
on scales of a few $\times 100$ $M_\odot$.

Various feedback effects, such as $\Htwo$ photodissociation from
internal UV fields and radiation pressure due to Ly$\alpha$ photon
trapping, are then likely to regulate the eventual efficiency of star
formation in these systems. These important questions can only be
addressed with some degree of confidence by high-resolution numerical
simulations that are able to track the detailed gas hydrodynamics,
chemistry and cooling, paralleling the pioneering work already done
for $T_{\rm vir} < 10^{4}$ K halos (Bromm et al. 1999; Abel et al. 2000).  

Finally, it is worth noting that during the initial contraction of the
gas in halos with $T_{\rm vir} > 10^{4}$ K a significant fraction of
the cooling radiation may be emitted in the Ly$\alpha$ line,
especially toward high redshifts, where the contracting gas has a low
metallicity. The cooling radiation may be significant, and can be
detectable for halos with circular velocities above $\sim 100$ ~km~s$^{-1}$,
as an extended, diffuse, low-surface brightness ``fuzz,'' with an
angular diameter of a few arcseconds (Haiman, Spaans, \& Quataert
2000; Fardal et al. 2001).  A quasar turning on during the early
stages of the contraction of the gas can boost the surface brightness
by a factor of $\sim 100$ (Haiman \& Rees 2001).

\section{Observational Prospects}

\subsection{Implications of Known High-Redshift Sources}

Observations over the last two years have uncovered a handful of
objects at redshifts around, and exceeding, $z=6$.  Quasars discovered
at $z\approx 6$ range from the very bright sources found in the Sloan
Digital Sky Survey (SDSS; $M_B\approx -27.7$ mag at $5.8\lsim z\lsim 6.4$; Fan
et al. 2000, 2001, 2003) to the much fainter quasars found using the
Keck telescope ($M_B=-22.7$ mag at $z=5.5$; Stern et al. 2000).  Galaxies
around the same redshifts are being discovered via their Ly$\alpha$
emission lines\footnote{This type of search was proposed over 30 years
ago by Partridge \& Peebles (1967).} (Dey et al. 1998; Spinrad et al. 1998; 
Weymann et al. 1998; Hu, McMahon, \& Cowie 1999), with some recent
extreme examples ranging from star formation rates of $>10-20\,
M_\odot {\rm yr^{-1}}$ at $z=5.7-6.6$ (Rhoads \& Malhotra 2001; Hu et
al. 2002; Kodaira et al. 2003; Rhoads et al. 2003), to an
exceptionally faint (with an inferred star formation rate of
$0.5\,M_\odot {\rm yr^{-1}}$) galaxy at $z=5.56$ detected in a
targeted search for gravitationally lensed and highly magnified
sources behind an Abell cluster (Ellis et al. 2001).

The spatial volume that was searched to discover the various sources
listed above spans $\sim 9$ orders of magnitude: the SDSS survey
probed $\sim 20$ Gpc$^3$ to discover the bright $z\approx  6$ quasars,
while the faint, strongly lensed Ly$\alpha$ galaxy was found in
searching a mere $\sim 100$ Mpc$^3$.  It is possible to associate these
high-redshift sources with dark matter halos based simply on their
inferred abundance: this suggests that the rare SDSS quasars reside in
massive, $\sim 10^{13}\,{M_\odot}$ halos, corresponding to
$4-5\sigma$ peaks in the primordial density field on these scales,
while the faintest Ly$\alpha$ galaxies may correspond to nearly
``$M_*$'' halos with masses of $\sim 10^{10}\,{M_\odot}$.  To
first approximation, the existence of these objects can be naturally
accommodated in the CDM structure formation models. We will next list
some simple conclusions that can be drawn from the existence of the
above sources.

\subsubsection{Early Growth of Massive Black Holes}

It is interesting to consider the sheer size of supermassive 
BHs required to power the bright SDSS quasars near $z\approx
6$.  Assuming that these quasars are shining at their Eddington limit,
and are not beamed or lensed\footnote{Strong lensing or beaming would
contradict the large proximity effect around these quasars; see Haiman
\& Cen (2002) and discussion below.}, their BH masses are inferred to
be $M_\bullet\approx 4\times10^9\,\msun$.  The Eddington-limited growth
of these supermassive BHs by gas accretion onto stellar-mass seed
holes, with a radiative efficiency of $\epsilon\equiv L/\dot
mc^2\approx 10\%$, requires $\sim 20$ $e$-foldings on a time scale of
$t_{\rm E}\approx 4\times10^7(\epsilon/0.1)$ years.  While the age of the
Universe leaves just enough time ($\lsim 10^9$ years) to accomplish
this growth by redshift $z=6$, it does mean that accretion has to
start early, and the seeds for the accretion have to be present at
ultra-high redshifts: $z\gsim 15$ or 20 for an initial seed mass of 100 or 10 
$\msun$, respectively. This conclusion holds even when one considers that BH
seeds may grow in parallel in many different early halos, which
undergo subsequent mergers (Haiman \& Loeb 2001). Furthermore, the
radiative efficiency cannot be much higher than $\epsilon\approx 10\%$
(Haiman \& Loeb 2001; Barkana, Haiman, \& Ostriker 2001).  Since an
individual quasar BH could have accreted exceptionally fast (exceeding
the Eddington limit), it will be important to apply this argument to a
larger sample of high-redshift quasars. Nevertheless, we note that a
comparison of the light output of quasars at the peak of their
activity ($z\approx 2.5$) and the total masses of their remnant BHs at
$z=0$ (see the review by Richstone, this volume) shows that
during the growth of most of the BH mass the radiative efficiency
cannot be much smaller than $10\%$, and hence any ``super-Eddington''
phase must be typically restricted to building only a small fraction
of the final BH mass (Yu \& Tremaine 2002).

\subsubsection{Amplification by Gravitational Lensing?}

A caveat to any conclusion based on the observed fluxes of bright,
distant quasars is that they may be gravitationally lensed, and
strongly amplified.  While the {\it a priori}\ probability of strong
lensing, causing amplification by a factor of $>10$, along a random
line of sight is known to be small ($\tau\approx 10^{-3}$) even to
high redshifts (e.g., Kochanek 1998; Barkana \& Loeb 2000), the {\it a
posteriori}\ probabilities for observed sources can be much higher due
to magnification bias (see, e.g., Schneider 1992).  Magnification bias
depends strongly on the parameters of the intrinsic quasar luminosity
function, which are poorly constrained at $z\approx 6$. As a result,
the theoretically expected probability that the SDSS quasars are
strongly amplified by lensing can be significant, even approaching
unity if the quasar luminosity function has an intrinsic slope steeper
than -dlog$\Phi$/dlog$L$$\gsim$4 and/or has a break at relatively
faint characteristic luminosities (Comerford, Haiman, \& Schaye 2002;
Wyithe \& Loeb 2002a,b).  As a result, observed lensed fractions can
be used to provide interesting constraints on the high-redshift quasar
luminosity function (Comerford et al. 2002; Fan et al. 2003).

Haiman \& Cen (2002) analyzed the flux distribution of the \lya
emission of the quasar with one of the highest known redshifts, SDSS
1030+0524 at $z=6.28$, and argued that this object could not have
been magnified by lensing by more than a factor of $\sim 5$.  The
constraint arises from the large observed size, $\sim$30 (comoving)
Mpc, of the ionized region around this quasar, and relies crucially
only on the assumption that the quasar is embedded in a largely
neutral intergalactic medium (IGM).  Based on the line/continuum ratio
of SDSS 1030+0524, this quasar is also unlikely to be beamed by a
significant factor.  The conclusion is that the minimum mass for its
resident BH is $4\times 10^8~\,{M_\odot}$ (for magnification by a
factor of 5); if the mass is this low, then the quasar had to switch
on prior to redshift $z_f\ga9$.  From the large size of the ionized
region, an absolute lower bound on the age of this quasar also follows
at $t>2\times 10^{7}~$yrs (see also the review by Martini on quasar
ages in this volume).

\subsubsection{Ly$\alpha$ Emitters and Cold Dark Matter}

The existence of the faintest Ly$\alpha$ emitters may have another
interesting implication.  Three faint sources were found by probing a
volume of only about $\sim$10 Mpc$^3$ (in a source plane area of
$\Delta \Omega \approx 100$ arcsec$^2$ behind the cluster Abell 2218 and
redshift range $\Delta z \approx 1$; Ellis et al. 2001). Associating the
implied spatial abundance of a few $\times$ 0.1 Mpc$^{-3}$ with those
of CDM halos (Jenkins et al. 2001), these sources correspond to very
low-mass ($M\approx 10^{10}\,{M_\odot}$) halos. This appears
consistent with the very low star formation rates ($\sim 0.5\,
M_\odot {\rm yr^{-1}}$), inferred from the Ly$\alpha$ luminosity.

The existence of such low-mass halos is interesting from the
perspective of other recent observations, which suggest that standard
CDM models predict too much power for the primordial density
fluctuations on small scales (see, e.g., Haiman, Barkana, \& Ostriker 
2001 for a brief review).  Several modifications of the CDM models, 
exemplified by warm dark matter (WDM) models (e.g., Bode, Ostriker, \& Turok 
2001), have been proposed recently that reduce the small-scale power.
Such modifications generally reduce the number of
low-mass halos at high redshift, and if the WDM particle had a mass
of $m_X \lsim 1$~keV (or $z=0$ velocity dispersion of $\upsilon_{\rm rms} \gsim
0.04$ ~km~s$^{-1}$), then there may have been too few high-$z$ sources to
reionize the Universe by $z=6$ (Barkana et al. 2001). 

The faint Ly$\alpha$-emitting galaxies are so far down on the mass
function of halos that one can turn this into a similar constraint on
the mass of the WDM particle.  Indeed, for $m_X \lsim 1$~keV, such low-mass 
halos would not exist at $z=6$ (see Fig. 5 in Barkana et al.
2001). This constraint is of interest, since it is around
the value of other current astrophysical limits (e.g., from the
Ly$\alpha$ forest; Narayanan et al. 2000).

\subsection{The Reionization History of the IGM}

How and when the intergalactic plasma was reionized is one of the 
long-outstanding questions in astrophysical cosmology, likely holding many
clues about the nature of the first generation of light sources and
the end of the cosmological ``Dark Age.''  The lack of any strong H~I
absorption (a GP trough) in the
spectra of high-redshift quasars has revealed that the IGM is highly
ionized at all redshifts $z \lsim 6$ (Fan et al. 2000).  On the other
hand, the lack of a strong damping by electron scattering of the first
acoustic peak in the temperature anisotropy of the CMB
radiation has shown that the IGM was neutral between
the redshifts $25\lsim z \lsim 10^3$ (Kaplinghat et al. 2003).
Together these two sets of data imply that most hydrogen atoms in the
Universe were reionized during the redshift interval $6\lsim z \lsim
25$.

It would be overly ambitious to provide a comprehensive review of the
subject of reionization in this article.  Instead, we will focus below
on a few basic theoretical issues and discuss the implications of the
most recent observations.

\subsubsection{Models of Reionization}

In the simplest models, an early population of stars or quasars drive
expanding, discrete ionized regions.  Once these regions overlap, the
Universe has been reionized (e.g., Arons \& Wingert 1972). In the
context of CDM models, the ionizing sources form inside high-redshift
dark matter halos. Since the formation and evolution of the halos is
dictated by gravity alone, it is relatively well understood, and the
main uncertainty in models of reionization is the ``efficiency''
(ionizing luminosity of stars and/or quasars) of each halo.
Semi-analytical models (e.g., Shapiro et al. 1994; Tegmark, Silk, \& Blanchard 
1994; Haiman \& Loeb 1997, 1998; Valageas \& Silk 1999) and
numerical simulations (Gnedin \& Ostriker 1997; Nakamoto, Umemura, \&
Susa 2001; Gnedin 2000, 2003) adopt various, reasonably motivated,
efficiencies and follow the evolution of the total volume-filling
fraction of ionized regions.  These models predict reionization to
occur between $z$ = 7 and 15, depending on the adopted efficiencies.

These studies have left significant uncertainties on the details of
how reionization proceeds in an inhomogeneous medium.  Since the
ionizing sources are likely embedded in dense regions, one might
expect that these dense regions are ionized first, before the
radiation escapes to ionize the low-density IGM (Madau, Haardt, \&
Rees 1999; Gnedin 2000, 2003).  Alternatively, most of the radiation
might escape from the local, dense regions along low-column density
lines of sight.  In this case, the underdense ``voids'' are ionized
first, with the ionization of the denser filaments and halos lagging
behind (Miralda-Escud\'e, Haehnelt, \& Rees 2000).

Apart from the topology of reionization, the current suite of models
also leaves uncertainties about the redshift evolution of the mean
ionized fraction.  Most models predict a sharp increase whenever the
discrete ionized regions percolate.  However, if the formation rate of
the ionizing sources does not parallel the collapse of high-$\sigma$
peaks, reionization can be more gradual, and can have a complex
history.  For instance, the radiative ${\rm H_2}$ feedback discussed
above may result in two distinct episodes of reionization:
(1) UV sources formed via ${\rm H_2}$ cooling in minihalos
partially ionize the Universe at $z\approx 20$, (2) the IGM recombines
as these sources turn off, and (3) the Universe is reionized at
$z\approx 7$ by UV sources in more massive halos. The first episode of
reionization may be more pronounced, since the metal-free stars in
minihalos are expected to have an unusually high ionizing photon
production efficiency (see also Wyithe \& Loeb 2003a; Cen 2003a).

Depending on the choice of the efficiency parameters, it is also
possible that the IGM only partially recombines during stage (2),
resulting in an extended episode of partial ionization (Cen 2003a).
Finally, the decrease of the mean neutral fraction would be more
gradual if the ionizing sources had a hard spectrum.  Reionization by
X-rays was considered recently by Oh (2001) and Venkatesan, Giroux, \&
Shull (2001).  In contrast to a picture in which discrete H~II regions
eventually overlap, in this case the IGM is ionized uniformly and
gradually throughout space.  All of these uncertainties highlight the
need for new and sensitive observational probes of the reionization
history, which we will discuss in the next section.

A significant source of theoretical uncertainty in the above models is
the average global recombination rate, or ``clumping factor,'' which
limits the growth of H~II regions at high redshifts, when the Universe
was dense.  Although hydrodynamical simulations can compute gas
clumping {\it ab initio}\ (Gnedin \& Ostriker 1997), to date they have
not been able to resolve the relevant small scales (the minihalos have
typical masses below $10^7\,\msun$).  Gas clumping has been estimated
semi-analytically (Chiu \& Ostriker 2000; Benson et al. 2001; Haiman,
Abel, \& Madau 2001).  In particular, Haiman et al. (2001) pointed out
that the earliest ionizing sources are likely surrounded by numerous
``minihalos'' that had collapsed earlier, but had failed to cool and
form any stars or quasars.  The mean-free path of ionizing photons,
before they are absorbed by a minihalo\footnote{Photoionization
unbinds the gas in these shallow potential wells (Shapiro, Raga, \&
Mellema 1998; Barkana \& Loeb 1999).  The gas acts as a sink of
ionizing photons only before it is photoevaporated.}, is about $\sim
1$ (comoving) Mpc.  Simple models, summing over the expected
population of minihalos, reveal that on average an H atom in the
Universe recombines $\gsim 10$ times before redshift $z=6$; as a
result, the IGM had to be ``reionized $\gsim 10$ times.''

A naive extrapolation of the luminosity density of bright quasars
toward $z=6$ reveals that these sources fall short of this
requirement (Haiman et al. 2001; see also Shapiro et al. 1994).
Extrapolating the known population of Lyman-break galaxies (e.g., Steidel, 
Pettini, \& Adelberger 2001) toward $z=6$ comes closer: assuming that 15$\%$ 
of the ionizing radiation from Lyman-break galaxies escapes into
the IGM (on average, relative to the escape fraction at 1500 \AA), a
naive extrapolation shows that Lyman-break galaxies emitted approximately 
one ionizing photon
per hydrogen atom prior to $z=6$. The implication is that the ionizing
emissivity at $z>6$ was $\sim10$ times higher than provided by a
straightforward extrapolation back in time of known quasar and galaxy
populations.  The Universe was likely reionized by a population of UV
sources that is yet to be discovered!  

\subsubsection{Current Observations}

The recent discovery of the bright quasar SDSS 1030+0524 in the 
SDSS at redshift $z=6.28$ has, for the first
time, revealed a full GP trough, i.e., a spectrum consistent with no
flux at high S/N over a substantial stretch of wavelength shortward of
$(1+z)\lambda_\alpha=8850$ \AA\ (Becker et al. 2001).  At the time of
this writing, a full GP trough has been discovered at high S/N in a
second SDSS quasar at $z=6.43$, and at a lower S/N in two other SDSS
quasars (at $z=6.23$ and $z=6.05$; Fan et al. 2003; an ``incomplete''
trough was also reported at high S/N in a $z=5.7$ source by Djorgovski
et al. 2001).

These discoveries have raised the tantalizing possibility that we are
detecting reionization occurring near redshift $z\approx 6.3$.  The lack
of any detectable flux indeed implies a strong lower limit, $x_{\rm
H}\gsim 0.01$, on the mean mass-weighted neutral fraction of the IGM
at $z\approx 6$ (Cen \& McDonald 2002; Fan et al. 2002; Lidz et al. 2002;
Pentericci et al. 2002).  Still, the evolution of the IGM opacity
inferred from quasar spectra does not directly reveal whether we
have probed the neutral era.  Nevertheless, comparisons with numerical
simulations of cosmological reionization (Cen \& McDonald 2002; Fan et
al. 2002; Gnedin 2003; Lidz et al. 2002; Razoumov et al. 2002),
together with the rapid rise toward high redshifts of the neutral
fractions inferred from a sample of high-redshift quasars from
$5.5\lsim z \lsim 6$ (Songaila \& Cowie 2002), suggest that the IGM is
likely neutral at $z\gsim6.5$.

While there may be a theoretical bias for reionization occurring close
to $z\approx 6.3$, it is possible that the reionization history was
nonmonotonic, and/or lasted over a considerably longer redshift
interval, as was discussed above.  An observational probe of the
redshift history of reionization would be invaluable in constraining
such scenarios, and to securely establish when the cosmic Dark Age
ended.  Below we consider prospects to probe the reionization history
in future observations.  Another interesting issue (not discussed
further below) raised by the recent GP trough detections is: What is
the best way to interpret quasar spectra?  Using simply the wavelength
extent of the ``dark'' region (without any detectable flux) in the
spectrum is not, by itself, generally sufficient to give a strong
constraint on the global topology of neutral versus ionized regions,
because of stochastic variations. Other statistical measures need to
be developed (Barkana 2002; Lidz et al. 2002; Nusser et al. 2002).

\subsubsection{Future Probes of Reionization}

\vspace{\baselineskip}
{\em CMB Polarization.} An alternative way of probing deeper into the
dark ages is the study of CMB anisotropies: the free electrons
produced by reionization scatter a few percent of the CMB photons.
Interesting results may be imminent from the ongoing CMB satellite
experiment {\it WMAP}.  The discussion below is based largely on the
results of Kaplinghat et al. (2003); for a detailed review, see Haiman
\& Knox (1999).

Without reionization, the ``primordial'' polarization signal at large
angles would be negligible. However, CMB photons scattering in a
reionized medium boost the polarization signal --- likely making it
measurable in the future.  CMB polarization anisotropy at large angles
is very sensitive to the optical depth to electron scattering, and the
future experiment {\it Planck}\ (and if the optical depth is large,
then {\it WMAP} as well), will have the power to discriminate between
different reionization histories even when they lead to the same
optical depth.

One of the advantages of studying the CMB is that it probes the
presence of free electrons, and can therefore detect a neutral
hydrogen fraction of $x_{\rm H}=0.1$ and $x_{\rm H}=10^{-3}$ with
nearly equal sensitivity.  Physically, the CMB and the GP trough
therefore probe two different stages of reionization.  The CMB is
sensitive to the initial phase when $x_{\rm H}$ first decreases below
unity and free electrons appear, say, at redshift $z_e$.  On the
other hand, the (hydrogen) GP trough is sensitive to the end phase,
when neutral hydrogen atoms finally disappear, say, at $z_{\rm H}$.
In most models, these two phases coincide to $\lsim 10$\% of the
Hubble time, such that $z_e\approx z_{\rm H}$.  However, as argued above,
one can conceive alternative theories in which the two phases are
separated by a large redshift interval, and $z_e\gg z_{\rm H}$.

One of the difficulties with CMB is that the effect of electron
scattering on the temperature anisotropies is essentially an overall
suppression, nearly degenerate with the intrinsic amplitude of the
fluctuation power spectrum. However, Kaplinghat et al. (2003) showed
that for most models constrained by current CMB data and by the
discovery of a GP trough (i.e., requiring that reionization occurred
at $z > 6.3$), {\it WMAP} can break this degeneracy, and detect the
reionization signature in the polarization power spectrum.\footnote{A
detailed morphological study of the effects of reionization on maps of
the temperature anisotropy may also be helpful to break this
degeneracy (Gnedin \& Shandarin 2002).}  The expected 1$\sigma$ error
on the measurement of the electron optical depth is around $\delta
\tau\approx 0.03$, with only a weak dependence on the actual value of
$\tau$.  This will also allow {\it WMAP} to achieve a 1$\sigma$ error
on the amplitude of the primordial power spectrum of 6\%. As an
example, {\it WMAP} with two years ({\it Planck}\ with one year) of
observation can distinguish a model with 50\% (6\%) partial ionization
between redshifts of 6.3 and 20 from a model in which hydrogen was
completely neutral at redshifts greater than 6.3.  {\it Planck}\ will
be able to distinguish between different reionization histories even
when they imply the same optical depth to electron scattering for the
CMB photons (Holder et al. 2003; Kaplinghat et al. 2003).

\vspace{\baselineskip}
{\em Ly$\alpha$ Emitters.} An alternative method to probe the
reionization history is to utilize the systematic changes in the
profiles of Ly$\alpha$ emission lines toward higher redshift. The
increased hydrogen IGM opacity beyond the reionization redshift makes
the emission lines appear systematically more asymmetric, and the
apparent line center systematically shifts toward longer
wavelengths, as absorption in the IGM becomes increasingly more
important and eliminates the blue side of the line (Haiman 2002;
Madau 2003a).  Because of the intrinsically noisy Ly$\alpha$ line
shapes, this method will require a survey that delivers a large sample
of Ly$\alpha$ emitters (Rhoads et al. 2002).

Ly$\alpha$ photons injected into a neutral IGM are strongly scattered,
and the red damping wing of the GP trough can strongly suppress, or
even completely eliminate, the Ly$\alpha$ emission line
(Miralda-Escud\'e 1998; Miralda-Escud\'e \& Rees 1998; Loeb \& Rybicki
1999).  Resonant absorption by the IGM may itself extend to the red
side of the line, if there is still significant cosmological gas
infall toward the source (Barkana \& Loeb 2003).  The reionization of
the IGM may therefore be accompanied by a rapid decline in the
observed space density of Ly$\alpha$ emitters beyond the reionization
redshift $z_r$ (Haiman \& Spaans 1999). Indeed, such a decline could
by itself provide a useful observational probe of the reionization
epoch in a large enough sample of Ly$\alpha$ emitters (Haiman \&
Spaans 1999; Rhoads \& Malhotra 2001), complementary to methods
utilizing the GP trough.

As shown by Cen \& Haiman (2000) and Madau \& Rees (2000), a source
with a bright ionizing continuum can create a large ($\gsim 30$
comoving Mpc) cosmological H~II region.  For a sufficiently luminous
source, and/or for a sufficiently wide intrinsic Ly$\alpha$ line width,
the size of the H~II region corresponds to a wavelength range
$\Delta\lambda$ that exceeds the width of the emission line, allowing
most of the intrinsic Ly$\alpha$ line to be transmitted without
significant scattering.  Furthermore, even for faint sources with
little ionizing continuum, a significant fraction of the emission line
can remain observable if the intrinsic line width is $\Delta \upsilon \gsim
300$ ~km~s$^{-1}$ (Haiman 2002).

%%%%%%%%%%%%%%%%%%%%%%%%% Figure 1 %%%%%%%%%%%%%%%%%%%%%%%%
\begin{figure}
\plotone{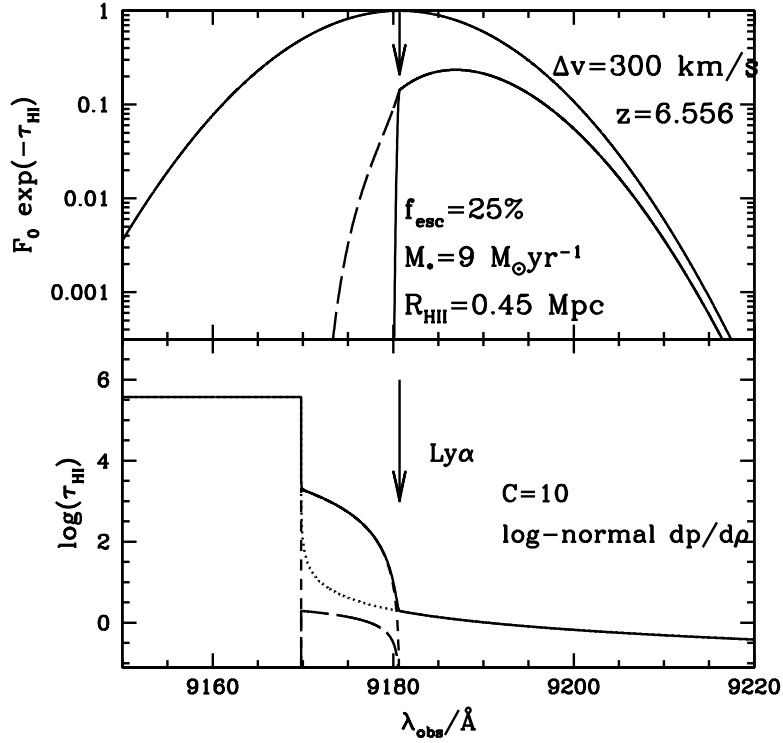}
\caption{{\it Upper panel:} Toy model profile of the \lya line from a
$z=6.56$ galaxy. The top solid curve shows the adopted intrinsic
profile, and the bottom solid curve shows the profile including
absorption in the IGM and by the neutral atoms inside the $0.45$
(proper) Mpc H~II region surrounding the source.  {\it Lower panel:}
The optical depth as a function of wavelength from within the H~II
region (short-dashed curve), from the neutral IGM outside the H~II
region (dotted curve), as well as from the sum of the two (solid
curve).  In both panels, the long-dashed curves describe an
alternative, more realistic treatment of the H~I opacity within the H~II
region (see Haiman 2002 for a discussion), and the arrows indicate the
central wavelength of the unobscured Ly$\alpha$ line.}
\label{fig:spec}
\end{figure}
%%%%%%%%%%%%%%%%%%%%%%%%%%%%%%%%%%%%%%%%%%%%%%%%%%%%%%%

The recent discovery of Ly$\alpha$-emitting galaxies with the Keck and
Subaru telescopes at redshifts as high as $z=6.56$ (Hu et al. 2002)
and $z=6.58$ (Kodaira et al. 2003) illustrates the fact the Ly$\alpha$
lines can indeed be detected even at these high redshifts. These
redshifts exceed those at which the GP troughs were discovered in the
SDSS quasar spectra, and hence these galaxies could be located {\it
beyond} the reionization redshift of the IGM.  The Hu et al. spectrum
is consistent with being embedded in a neutral IGM that only partially
obscures the line, and allows $\sim20\%$ of the total line flux to be
transmitted (Haiman 2002; see Fig.~\ref{fig:spec}).  This would imply
that the source is somewhat surprisingly bright, given the inferred
abundance and estimates of the number density of $z\approx 5.7$
Ly$\alpha$ emitters from a different survey (Rhoads et al. 2003).

Although the lines are detectable, the IGM still has a significant
effect on the Ly$\alpha$ line profile.  A statistical sample of
Ly$\alpha$ emitters that spans the reionization redshift should be a
useful probe of reionization, through the study of the correlations
between the luminosity of the sources and the properties of the
emission lines, such as their total line/continuum ratio (if a
continuum is measured), the asymmetry of the line profile, and the
offset of the peak of the line from the central Ly$\alpha$ wavelength
(for sources that have redshift measurements from other emission
lines).

\vspace{\baselineskip} {\em Redshifted 21 cm Features.} Future radio
telescopes could observe 21 cm emission or absorption from neutral
hydrogen at the time of reionization (see a recent review by Madau
2003b). This would provide a direct measure of the physical state of
the neutral hydrogen and its evolution through the time of
reionization.  Recently, Carilli, Gnedin, \& Owen (2002) considered
the radio equivalent of the GP trough, using numerical
simulations. Unlike the Ly$\alpha$ case, the mean absorption by the
neutral medium is about 1\% at the redshifted 21 cm.  Furlanetto \&
Loeb (2002) and Iliev et al. (2002) have used semi-analytic methods to
look at the observable features (in 21 cm) of minihalos and
protogalactic disks.  These studies suggest that the 21 cm
observations would yield robust information about the thermal history
of collapsed structures and the ionizing background, provided that a
sufficiently bright radio-loud quasar can be found at $z_e > 6.3$. In
addition, characteristic angular fluctuations that trace early density
fluctuations of the 21-cm emitting gas (e.g., Madau, Meiksin, \& Rees
1997; Tozzi et al. 2000) may be detectable in the future.

\vspace{\baselineskip} 
{\em Gunn-Peterson Trough in Metal Lines.}
Although detections of the hydrogen GP trough suffer from the
``saturation problem'' discussed above, an alternative possibility may
be to use corresponding absorption troughs caused by heavy elements in
the high-redshift IGM.  Recently, Oh (2002) showed that if the IGM is
uniformly enriched by metals to a level of $Z=10^{-2}-10^{-3}~{Z_\odot}$, 
then absorption by resonant lines of O~I or Si~II may be
detectable.  The success of this method depends on the presence of
oxygen and silicon at these levels, and on these species being
neutral or once-ionized in regions where hydrogen has not yet been
ionized. It may be more natural for hydrogen reionization to
precede metal enrichment, rather than vice versa, however; because of
the short recombination time at high redshift, the gas can remain 
neutral even in the metal-enriched regions, and may provide the 
absorption features necessary for its detection.

\vspace{0.3cm}
{\bf Acknowledgements}.
I thank Carnegie Observatories for their kind invitation, and
my recent collaborators Gil Holder, Manoj Kaplinghat, Lloyd Knox, and
Peng Oh for many fruitful discussions, and for their permission to
draw on joint work.

\section*{Appendix: Implications of First WMAP Results}
\label{sec:appendix}

%%%%%%% Figure 2 %%%%%%
\begin{figure}[t]
\begin{center}
\plotone{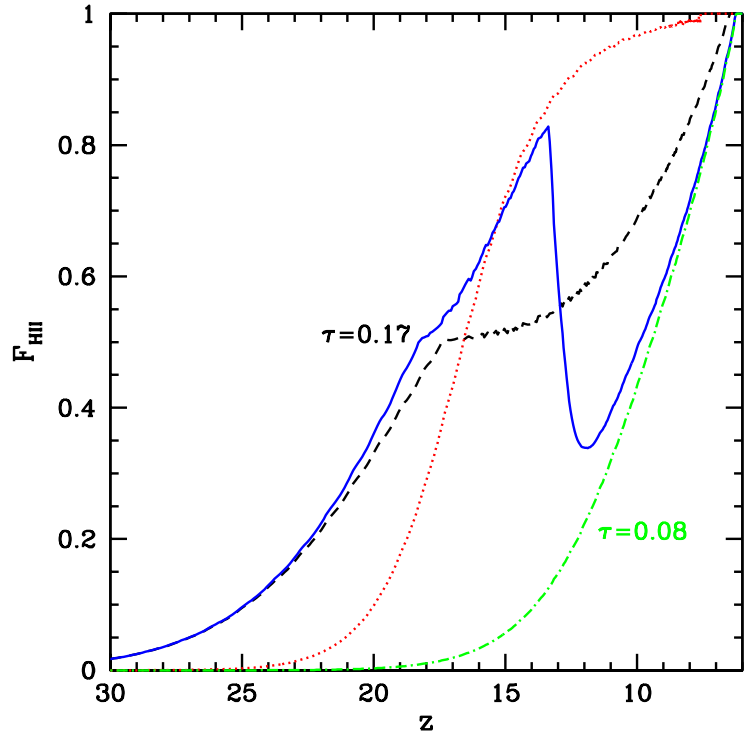}
\end{center}
\caption[]{Model reionization histories (evolution of ionized fraction
with redshift).  The dot--dashed curve is a simple model with
``universal'' efficiencies calibrated to produce reionization at
$z=6$. It produces a low optical depth of $\tau=0.08$.  The other
three models satisfy the dual requirement of (1) a low--redshift
($z\sim 6$) percolation epoch, and a (2) high value of electron
scattering optical depth ($\tau\sim 0.17$) by (1) increased efficiency
in ``mini--halos'' (solid curve), (2) excluding minihalos, but
increasing the efficiencies in larger halos (dotted curve), and (3) a
sudden drop in efficiencies, (e.g. due to a transition from metal-free
to normal stellar population).  The three models produce large--angle
polarization signatures in the CMB that will be distinguishable by
Planck (adopted from Haiman \& Holder 2003).  }
\label{fig:reion}
\end{figure}
%%%%%%%%%%%%%%%%%%%%%%%%

As this article went to press, the first results by the {\it WMAP}\
experiment were announced.  These new results have significant
implications for reionization, which we only briefly summarize here.
{\it WMAP} has measured the optical depth to electron scattering from
the cross-correlation between the the temperature and $E$-mode
polarization angular power spectra (TE; Spergel et al. 2003), yielding
the high value of $\tau=0.17\pm0.06$.  The value reflects
marginalization over all other relevant cosmological parameters.  A
reassuringly similar value, $\tau=0.17\pm0.04$, is obtained by
predicting the TE power spectrum directly from the temperature power
spectrum (Kogut et al. 2003).

The immediate conclusion that can be drawn from the high value of
$\tau$ is that the IGM was significantly ionized at redshifts as early
as $z\approx 15$.  This discovery has important implications for the
sources of reionization, and allows, for the first time, constraints
to be placed on physical reionization scenarios out to redshift $z\approx 
20$.  The {\it WMAP} results have been interpreted by a flurry of
papers presenting models of the reionization history in the few weeks
following the data release (Cen 2003b; Ciardi, Ferrara, \& White 2003; 
Fukugita \& Kawasaki 2003; Haiman \& Holder 2003; Sokasian, Abel, \& Hernquist 
2003; Somerville \& Livio 2003; Wyithe \& Loeb 2003b). The reader is
referred to these papers for many interesting details; no doubt the
list of papers devoted to reionization will continue to grow by the
time this article appears in print.

The main implications of the {\it WMAP} result can be summarized as
follows (based on Haiman \& Holder 2003, but broadly consistent with
the conclusions reached in all the interpretive papers listed above).

%\begin{itemize} 

\vspace{\baselineskip} \noindent$\bullet$ Previous evidence has shown
that the IGM is highly ionized at least out to redshift $z\approx 6$.  As
argued in the body of this article, there is also evidence that a
``percolation epoch'' is taking place near $z\approx 6-7$.\footnote{Hui
\& Haiman (2003) have argued that there is additional evidence for
percolation at $z<10$ from the {\it thermal} history of the IGM.}
Abrupt reionization at $z=6$ would yield $\tau\approx 0.04$,
significantly lower than the {\it WMAP} value.

\vspace{\baselineskip} \noindent$\bullet$ Reionization models predict
a ``percolation'' redshift that depends on the combination of
efficiency parameters, essentially on $\epsilon \equiv N_\gamma f_*
f_{\rm esc}/C$, where $f_* \equiv M_*/(\Omega_{\rm b}M_{\rm
halo}/\Omega_{\rm m})$ is the fraction of baryons in the halo that
turns into stars; $N_\gamma$ is the mean number of ionizing photons
produced by an atom cycled through stars, averaged over the initial
mass function of the stars; $f_{\rm esc}$ is the fraction of these
ionizing photons that escapes into the IGM; and $C$ is the mean
clumping factor of ionized gas.  A value of $\epsilon\approx 10$ is
required for percolation to occur at $z\approx 6$.  This value is
quite reasonable, and it produces a natural ``tail'' of ionization at
redshifts exceeding $z=6$ in these models (due to the gradual turn-on
of ionizing sources associated with dark halos; see
Figure~\ref{fig:reion}). This ``tail'' increases the optical depth to
$\tau\approx 0.08$, which is still discrepant with the {\it WMAP}
value at the $\sim 3\sigma$ level.

\vspace{\baselineskip} \noindent$\bullet$ As a result, no simple
reionization model can be consistent with the combination of the
central {\it WMAP} value of $\tau=0.17$ {\it and} a percolation
occurring at $z\approx 6$ .  Satisfying both constraints requires
either of the following: (1) ${\rm H_2}$ molecules form efficiently at
$z\approx 20$, survive feedback processes, and allow UV sources in
halos with virial temperatures $T_{\rm vir}<10^4$ K to contribute
substantially to reionization, or (2) the efficiency $\epsilon$ in
halos with $T_{\rm vir}>10^4$ K decreased by a factor of $\gsim 30$
between $z\approx 20$ and $z\approx 6$.  The latter may be a natural
result of a switch-over from a metal-free to a normal stellar
population (Wyithe \& Loeb 2003a; Cen 2003a).  These options are
illustrated by the upper three curves in Figure~\ref{fig:reion}.

\vspace{\baselineskip} \noindent$\bullet$ As apparent from above,
there are interesting implications for the formation history of
ionizing sources, but there is no ``crisis'' for cosmology:
$\Lambda$CDM cosmogonies can still accommodate the high value of
$\tau$ measured by {\it WMAP}. However, interesting limits can be
drawn on cosmological models with reduced small-scale power.  As an
example, the combination of {\it WMAP} and other large-scale
structure data has provided tentative evidence for a running spectral
index in the power spectrum $P(k)$.  Haiman \& Holder (2003) and
Somerville et al. (2003) have shown that in the favored running-index
model achieving $\tau=0.17$ is impossible without extreme
efficiencies of ionizing photon production in halos with virial
temperatures $T_{\rm vir}<10^4$ K; a significantly stronger curvatures
could be ruled out.  Similar conclusions can be drawn (Somerville, Bullock, 
\& Livio 2003) about models with a strong tilt in the scalar power-law
index $n$, or about WDM models (see also Barkana et al.  2001 and Yoshida et 
al. 2003 for limits even on ``lukewarm'' dark matter models from reionization 
at $z\approx 20$.)

%\end{itemize} 

\vspace{\baselineskip} It is also worth emphasizing that the
reionization history is likely to have been complex enough so that it
will have distinctive features that {\it Planck}\ can distinguish at
$> 3\sigma$ significance (see Fig.~\ref{fig:reion}; and also Holder et
al. 2003).  At the high {\it WMAP} value for $\tau$, {\it Planck}\
will be able to provide tight statistical constraints on reionization
model parameters and help elucidate the nature of the sources ending
the Dark Ages.  In addition to the large-angle polarization signatures
from reionization, small-scale fluctuations in the IGM temperature may
be observable (a ``Sunyaev-Zel'dovich'' effect from high-redshift
ionized regions; Oh, Cooray, \& Kamionkowski 2003).  Finally, the
sources responsible for the high optical depth discovered by {\it
WMAP} should be directly detectable out to $z\approx 15$ by the {\it
JWST}.

\end{document}